\documentclass[aps,prc]{revtex4}
\begin{document}
\title{Jastrow-type calculations of one-nucleon removal reactions on open
$s$-$d$ shell nuclei}
\author{M.K. Gaidarov, K.A. Pavlova, and A.N. Antonov}
\affiliation{Institute of Nuclear Research and Nuclear Energy,
Bulgarian Academy of Sciences, Sofia 1784, Bulgaria}
\author{C. Giusti}
\affiliation{Dipartimento di Fisica Nucleare e Teorica,
Universit\`a di Pavia,\\
Istituto Nazionale di Fisica Nucleare, Sezione di Pavia, Pavia,
Italy}
\author{S.E. Massen and Ch.C. Moustakidis}
\affiliation{Department of Theoretical Physics, Aristotle
University of Thessaloniki, GR-54006 Thessaloniki, Greece}
\author{K. Spasova}
\affiliation{Department of Theoretical and Applied Physics,
"Bishop K. Preslavski" University, Shumen 9712, Bulgaria}

\begin{abstract}
Single-particle overlap functions and spectroscopic factors are
calculated on the basis of Jastrow-type one-body density matrices
of open-shell nuclei constructed by using a factor cluster
expansion. The calculations use the relationship between the
overlap functions corresponding to bound states of the
$(A-1)$-particle system and the one-body density matrix for the
ground state of the $A$-particle system. In this work we extend
our previous analyses of reactions on closed-shell nuclei by using
the resulting overlap functions for the description of the cross
sections of $(p,d)$ reactions on the open $s$-$d$ shell nuclei
$^{24}$Mg, $^{28}$Si and $^{32}$S and of $^{32}$S$(e,e^{\prime}p)$
reaction. The relative role of both shell structure and
short-range correlations incorporated in the correlation approach
on the spectroscopic factors and the reaction cross sections is
pointed out.
\end{abstract}
\maketitle

\section{Introduction}
Nowadays consistent efforts in the correct treatment of the beyond
mean-field nucleon-nucleon (NN) correlations when performing
nuclear calculations have been made (see, e.g.,
\cite{Ant88,Ant93}). While collective phenomena have been known
since the early times of nuclear physics, effects produced by
short-range correlations (SRC) are more difficult to be
experimentally singled out. The experimental and theoretical works
in the past decade have provided us with a much clearer picture on
the consequences of these correlations. There are clear signatures
of the presence of correlation effects on some quantities related
to the behaviour of the single nucleon in nuclear medium. For
instance, NN correlations are responsible for the reduction of the
spectroscopic strengths of the nuclear hole states
\cite{Ant88,Ant93,Lap93}. The theoretical understanding of this
fact requires the knowledge of the removal spectral function which
can be represented in a natural way by both overlap functions and
single-nucleon spectroscopic factors \cite{An95}. They are
directly related to observable quantities such as $(p,d)$,
$(e,e^{\prime}p)$ and $(\gamma,p)$ reaction cross sections. The
comparison between the theoretically calculated and measured cross
sections could serve as a test for the proper account of
correlations within the theoretical approach considered.

Recently, a general procedure has been adopted \cite{Vn93} to
extract the bound-state overlap functions and the associated
spectroscopic factors and separation energies on the base of the
ground-state one-body density matrix (OBDM). Initially, the
procedure has been applied \cite{Sto96} to a model OBDM
\cite{Sto93} accounting for SRC within the low-order approximation
(LOA) to the Jastrow correlation method \cite{Jas55}. First, the
applicability of the theoretically calculated overlap functions
has been tested in the description of the $^{16}$O$(p,d)$ pickup
reaction \cite{Di97,Gai99} and of the $^{40}$Ca$(p,d)$ reaction
\cite{Di97,Iva2001}. {\it The main result} from these
investigations is that {\it absolute cross sections of the $(p,d)$
reactions are evaluated without any additional normalization in
contrast to standard Distorted Wave Born Approximation (DWBA)
calculations}. Then, a detailed study of the cross sections of the
electron- and photon-induced knockout reactions on $^{16}$O
\cite{Gai2000} and $^{40}$Ca \cite{Iva2001} has been performed.
The $(e,e^{\prime}p)$ and $(\gamma,p)$ reactions are more suitable
to test various single-particle (SP) overlap functions and NN
correlations. The theoretical results obtained in Refs.
\cite{Iva2001,Gai2000} reproduce with a fair agreement the shape
of the experimental cross sections, in particular at large values
of the missing momentum, where correlation effects are more
sizable. Of course, the general success of the above procedure
depends strongly on the availability of realistic one-body density
matrices.

In the various approaches, the OBDM's are usually constructed for
closed-shell nuclei. The $^{16}$O nucleus has been studied by
variational Monte Carlo \cite{Pie92} and Green function
\cite{Di92,Po96} methods. The treatment of heavier nuclei has not
yet attained the same degree of accuracy as the light ones. The
correlated basis function (CBF) theory, based on the Jastrow
approach, has recently been extended to medium-heavy doubly
closed-shell nuclei using Fermi hypernetted chain integral
equations \cite{Co92,Sa96}. In addition, $^{16}$O and $^{40}$Ca
nuclei have been examined by the generator coordinate method
\cite{Iva2000}.

There is no systematic study of the one-body density matrix (and
related quantities) which includes both closed- and open-shell
nuclei. For that reason, in \cite{Mous2000} expressions for the
OBDM's $\rho({\bf r},{\bf r^{\prime}})$ and momentum distributions
which could be used for both closed- and open-shell nuclei have
been obtained. In \cite{Mas99} analytical expressions for the
charge form factors and densities of the $s$-$p$ and $s$-$d$ shell
nuclei have been also derived. The $\rho({\bf r},{\bf
r^{\prime}})$ was constructed using the factor cluster expansion
of Clark and co-workers \cite{Cla70,Ris71,Cla79} and Jastrow
correlation function which incorporates SRC for closed-shell
nuclei. This approach was extrapolated to the case of $N$=$Z$
open-shell nuclei in Ref. \cite{Mous2000}. Analyzing three
different expansions it has been shown in \cite{Mous2001} that
they lead to almost equivalent results for important nuclear
properties. The expressions for $\rho({\bf r},{\bf r^{\prime}})$
are functionals of the spherical harmonic oscillator (HO) orbitals
and depend on the HO parameter and the correlation parameter. The
values of the parameters which have been used for the closed-shell
nuclei $^{4}$He, $^{16}$O and $^{40}$Ca were determined in
\cite{Mas99} by a fit of the theoretical charge form factor,
derived with the same cluster expansion, to the experimental one.
For the open-shell nuclei $^{12}$C, $^{24}$Mg, $^{28}$Si and
$^{32}$S new values of these parameters have been obtained to give
a better fit to the experimental data \cite{Mous2000}. Moreover,
these nuclei were treated as $1d$ shell nuclei and as $1d$-$2s$
shell nuclei. In the latter case the $A$-dependence of the
high-momentum components of the momentum distributions becomes
quite small.

The aim of the present work is to study the bound-state overlap
functions for open-shell nuclei on the basis of Jastrow-type
one-body density matrices for such nuclei. Next, the resulting
overlap functions are tested in the description of nuclear
observables, namely the cross sections of $(p,d)$ reactions on
$^{24}$Mg \cite{Kal75}, $^{28}$Si \cite{Sun69} and $^{32}$S
\cite{Kal75} and of the $^{32}$S$(e,e^{\prime}p)$ reaction
\cite{Wes92}. Such an investigation allows us to examine the
relationship between the OBDM and the associated overlap functions
in that region of nuclei and also to estimate the role of SRC
incorporated in the correlation approach used in the reaction
cross section calculations.

The paper is organized as follows. A short description of the
correlated OBDM of the target nucleus is given in Sec. II together
with the procedure to extract the SP overlap functions from it.
The results of the calculations are presented and discussed in
Sec. III. The summary of the present work is given in Sec. IV.

\section{Theoretical approach}
\subsection{Correlated one-body density matrix}
In order to evaluate the SP bound-state overlap functions, one
needs the ground-state OBDM of the target nucleus which is
defined by the expression:
\begin{equation}
\rho ({\bf r},{\bf r^{\prime} })=A \int \Psi ^{*}( {\bf r},{\bf
r}_{2},...,{\bf r}_{A}) \Psi ({\bf r^{\prime} },{\bf
r}_2,...,{\bf r}_{A}) d{\bf r}_{2}...d{\bf r}_{A},
\label{eq:OBDM}
\end{equation}
where $\Psi ({\bf r}_{1},{\bf r}_{2},...,{\bf r}_{A})$ is the
normalized $A$-nucleon ground-state wave function. The integration
in Eq. (\ref{eq:OBDM}) is carried out over the radius vectors
${\bf r}_{2}$ to ${\bf r}_{A}$ and summation over spin variables
is implied. In the present work we start from a Jastrow-type trial
many-particle wave function
\begin{equation}
\Psi ({\bf r}_{1},...,{\bf r}_{A})={\cal F}\Phi=
\prod_{i<j}^{A}f(r_{ij})\Phi ({\bf r}_{1},...,{\bf r}_{A}),
\label{eq:WF}
\end{equation}
where ${\cal F}$ is a model operator which introduces SRC, $\Phi $
is an uncorrelated (Slater determinant) wave function built up
from single-particle wave functions which correspond to the
occupied states, and $f(r_{ij})$ is the state-independent
correlation function which was chosen to have the form
\begin{equation}
f(r_{ij})=1-\exp[-\beta ({\bf r}_{i}-{\bf r}_{j})^{2}].
\label{eq:cf}
\end{equation}
The correlation function goes to 1 for large values of
$r_{ij}=|{\bf r}_{i}-{\bf r}_{j}|$ and to 0 for $r_{ij}\rightarrow
0$. Apparently, the effect of SRC introduced by the function
$f(r_{ij})$ becomes larger when the correlation parameter $\beta$
becomes smaller and vice versa.

For our aims we represent the one-body density matrix $\rho({\bf
r},{\bf r^{\prime}})$ by the form
\begin{equation}
\rho({\bf r},{\bf r^{\prime}})=\frac{\langle \Psi |{\bf O}_{{\bf
rr^{\prime}}}|\Psi^{\prime} \rangle
}{\langle\Psi|\Psi\rangle}=N\langle\Psi|{\bf O}_{{\bf
rr^{\prime}}}|\Psi^{\prime}\rangle=N\langle{\bf O}_{{\bf
rr^{\prime}}}\rangle,
\label{eq:newOBDM}
\end{equation}
where $\Psi^{\prime}=\Psi({\bf r}_{1}^{\prime},{\bf
r}_{2}^{\prime},...,{\bf r}_{A}^{\prime})$, $N$ is the
normalization factor and the integration is carried out over the
vectors ${\bf r}_{1}$ to ${\bf r}_{A}$ and ${\bf
r}_{1}^{\prime}$  to ${\bf r}_{A}^{\prime}$. The one-body
"density operator" ${\bf O}_{{\bf rr^{\prime}}}$ has the form
\begin{equation}
{\bf O}_{{\bf rr^{\prime}}}=\sum_{i=1}^{A} \delta({\bf r}_{i}-{\bf
r})\delta({\bf r}_{i}^{\prime}-{\bf r}^{\prime})\prod_{j\neq
i}^{A}\delta({\bf r}_{j}-{\bf r}_{j}^{\prime}).
\label{eq:doper}
\end{equation}
The same one-body operator has been used also in \cite{Mor99} [Eq.
(14) of this reference] for realistic study of the nuclear
transparency and the distorted momentum distributions in the
semi-inclusive process $^{4}$He$(e,e^{\prime}p)$X.

In order to evaluate the correlated one-body density matrix
$\rho_{cor}({\bf r},{\bf r^{\prime}})$ we consider firstly the
generalized integral
\begin{equation}
I(\alpha)=\langle\Psi|\exp\left[\alpha I(0){\bf O}_{{\bf
rr^{\prime}}}\right]|\Psi^{\prime}\rangle , \label{eq:integral}
\end{equation}
corresponding to the one-body "density operator" ${\bf O}_{{\bf
rr^{\prime}}}$ (given by (\ref{eq:doper})), from which we have
\begin{equation}
\langle{\bf O}_{{\bf rr^{\prime}}}\rangle=\left[\frac{\partial \ln
I(\alpha)}{\partial \alpha}\right]_{\alpha=0}. \label{eq:deriv}
\end{equation}
For the cluster analysis of Eq. (\ref{eq:deriv}), after
considering the sub-product integrals \cite{Cla70,Ris71,Cla79} and
their factor cluster decomposition following the procedure of
Ristig, Ter Low, and Clark, one can obtain an expression for the
$\rho_{cor}({\bf r},{\bf r^{\prime}})$ \cite{Mous2000}
\begin{equation}
\rho_{cor}({\bf r},{\bf r^{\prime}})\simeq N [\langle {\bf
O}_{{\bf rr^{\prime}}}\rangle_{1}-O_{22}({\bf r},{\bf
r^{\prime}},{\rm g_1})-O_{22}({\bf r},{\bf r^{\prime}},{\rm
g_2})+O_{22}({\bf r},{\bf r^{\prime}},{\rm g_3})].
\label{eq:clusOBDM}
\end{equation}

The one-body contribution to the OBDM, $\langle {\bf O}_{\bf
rr'}\rangle_1$ and the three terms $O_{22}({\bf r},{\bf r'},{\rm
g}_l)$ ($l=1,2,3$), which come from the two-body contribution,
have the general forms
\begin{equation}
\langle {\bf O}_{\bf rr'}\rangle_1=\rho_{SD}({\bf r},{\bf r}') =
\frac{1}{\pi} \sum_{nl} \eta_{nl} (2l+1)
 \phi^{*}_{nl}(r) \phi_{nl}(r') P_l(\cos \omega_{rr'} ),
\label{O1-1}
\end{equation}
and
\begin{eqnarray}
O_{22}({\bf r},{\bf r}',  {\rm g}_l)& = & 4 \sum_{n_i l_i,n_j l_j}
\eta_{n_i l_i} \eta_{n_j l_j} (2 l_i +1) (2 l_j +1 )  \nonumber \\
&  &\times \left[ 4 A_{n_il_in_jl_j}^{n_il_i n_jl_j,0 } ({\bf
r},{\bf r}', {\rm g}_l) - \sum_{k=0}^{l_i +l_j} \langle l_i 0 l_j
0 \mid k 0 \rangle^2
 A_{n_il_in_jl_j}^{n_jl_j n_il_i,k}({\bf r},{\bf r}',{\rm g}_l) \right],
\label{O22-g-3}
\end{eqnarray}
where $\eta_{nl}$ are the occupation probabilities of the various
states, $\phi_{nl}(r)$ are the radial SP wave functions and
\begin{eqnarray}
A_{n_1l_1n_2l_2}^{n_3l_3n_4l_4,k}({\bf r},{\bf r}', {\rm g}_1)& =&
\frac{1}{4\pi}\phi^{*}_{n_1l_1}(r) \ \phi_{n_3l_3}(r') \
\exp[-\beta r^2] \ P_{l_3}(\cos\omega_{rr'})
 \nonumber \\
& &\times \int_{0}^{\infty}\phi^{*}_{n_2l_2}(r_2)
\phi_{n_4l_4}(r_2) \exp[-\beta r_{2}^2] \ i_k (2 \beta r r_2)
 r_{2}^{2}   d r_2  .
\label{A-O22-1}
\end{eqnarray}
The matrix elements $A_{n_1l_1n_2l_2}^{n_3l_3n_4l_4,k}({\bf
r},{\bf r}', {\rm g}_l)$ for $l=2$ and $3$ have similar structure
and are given in Ref. \cite{Mous2000}. In the above expression
$\omega_{rr'}$ is the angle between the vectors ${\bf r}$ and
${\bf r'}$ and $i_k(z)$ the modified spherical Bessel function.

Expansion (\ref{eq:clusOBDM}) of the present work has one- and
two-body terms and the normalization of the wave function is
preserved by the normalization factor $N$. The expressions of
$\langle {\bf O}_{\bf rr'}\rangle_1$ and the three two-body terms
$O_{22}({\bf r},{\bf r'},{\rm g}_l)$ ($l=1,2,3$) given by Eqs.
(\ref{O1-1}) and (\ref{O22-g-3}) depend on the SP radial wave
functions and so they are suitable to be used for analytical
calculations with HO orbitals. These expressions were derived for
the closed-shell nuclei with $N=Z$, where $\eta_{nl}$ is $0$ or
$1$. For the open-shell nuclei (with $N=Z$) we use the same
expressions as in Refs. \cite{Mous2000} and \cite{Mas99} where now
$0\le \eta_{nl}\le 1$. In this way the systematic study of
important quantities such as bound-state overlap functions and
spectroscopic factors \cite{Di97,Gai99,Iva2001,Gai2000} can be
extended to open-shell nuclei within the factor cluster expansion
from Refs. \cite{Mous2000,Cla70,Ris71,Cla79}.

It should be noted that a similar expression for $\rho_{cor}({\bf
r},{\bf r^{\prime}})$ was derived by Gaudin et al. \cite{Gau71} in
the framework of the LOA mentioned in the Introduction. Their
expansion contains one- and two-body terms and a part of the
three-body terms so that the normalization property of the OBDM is
fulfilled.

\subsection{Overlap functions and their relation to the OBDM}

The quantities related to the $(A-1)$-particle system, such as
the overlap functions, the separation energies and the
spectroscopic factors for its bound states can be fully
determined in principle by the one-body density matrix for the
ground state of the $A$-particle system \cite{Vn93}. This unique
relationship holds generally for quantum many-body systems with
sufficiently short-range forces between the particles and is
based on the exact representation of the ground-state one-body
density matrix \cite{Vn93}. The latter can be expressed in terms
of the overlap functions $\phi_{\alpha}({\bf r})$, which describe
the residual nucleus as a hole state in the target, in the form:
\begin{equation}
\rho ({\mbox{\boldmath $r$}},{\mbox{\boldmath $r$}}^{\prime})=
\sum_{\alpha }\phi_{\alpha}^{*} ({\mbox{\boldmath
$r$}})\phi_{\alpha}({\mbox{\boldmath $r$}}^{\prime}).
\label{eq:ovf}
\end{equation}

In the case of a target nucleus with $J^{\pi }=0^{+}$ each of the
bound-state overlap functions is characterized by the set of
quantum numbers $\alpha \equiv nlj$, with $n$ being the number of
the state with a given multipolarity $l$ and a total angular
momentum $j$. The asymptotic behaviour of the radial part of the
neutron overlap functions for the bound states of the
$(A-1)$-system is given by \cite{Vn93,Bang85}:
\begin{equation}
\phi_{nlj}(r)\rightarrow C_{nlj}\exp(-k_{nlj}r)/r,
\label{eq:nasym}
\end{equation}
where
\begin{equation}
k_{nlj}=\frac{1}{\hbar} [2m_{\mathrm
n}(E_{nlj}^{A-1}-E_{0}^{A})]^{1/2}.
\label{eq:decay}
\end{equation}
In Eq. (\ref{eq:decay}) $m_{\mathrm n}$ is the neutron mass,
$E_{0}^{A}$ is the ground state energy of the target $A$-nucleus
and $E_{nlj}^{A-1}$ is the energy of the $nlj$-state of the
$(A-1)$-nucleus. For protons some mathematical complications arise
due to an additional long-range part of the interaction
originating from the Coulomb force, but the general conclusions of
the consideration remain the same. The asymptotic behaviour of the
radial part of the corresponding proton overlap functions reads
\begin{equation}
\phi_{nlj}(r)\rightarrow C_{nlj}\exp[-k_{nlj} r-\eta \ln
(2k_{nlj}r)]/r,
\label{eq:pasym}
\end{equation}
where $\eta $ is the Coulomb (or Sommerfeld) parameter and
$k_{nlj}$ (\ref{eq:decay}) contains the mass of the proton.

The lowest $n=n_{0}$ neutron $lj$-bound-state overlap function is
determined by the asymptotic behaviour of the associated partial
radial contribution of the one-body density matrix
$\rho_{lj}(r,r^{\prime})$ ($r^{\prime}=a\rightarrow \infty $) and
Eqs. (\ref{eq:ovf}) and (\ref{eq:nasym}) lead to the expression
\begin{equation}
\phi_{n_{0}lj}(r)={\frac{{\rho _{lj}(r,a)}}{{C_{n_{0}lj}~\exp
(-k_{n_{0}lj}\,a})/a}},
\label{eq:lowest}
\end{equation}
where the constants ${C_{n_{0}lj}}$ and ${k_{n_{0}lj}}$ are
completely determined by $\rho_{lj}(a,a)$. In this way the
separation energy
\begin{equation}
\epsilon _{n_{0}lj}\equiv
E_{n_{0}lj}^{A-1}~-~E_{0}^{A}=\frac{\hbar
^{2}~k_{n_{0}lj}^{2}}{2m_{\mathrm n}}
\label{eq:energy}
\end{equation}
and the spectroscopic factor $S_{n_{0}lj}=\langle
\phi_{n_{0}lj}|\phi_{n_{0}lj}\rangle$ can be determined as well.

The applicability of this theoretical scheme has been demonstrated
in Refs. \cite{Sto96,Di97,Gai99,Iva2001,Gai2000} by means of
realistic one-body density matrices of $^{16}$O and $^{40}$Ca
constructed within different correlation methods. In particular,
the calculated overlap functions corresponding to the OBDM's from
various approaches to the CBF theory \cite{Sa96,Vn97} have shown
the substantial deviation of their shapes in respect to the
Hartree-Fock (HF) wave functions \cite{Gai99}. The inclusion of
short-range and tensor correlations has caused a depletion of the
levels below Fermi level which is reflected in the substantial
reduction of the spectroscopic factors for the quasihole states.

Thus having the procedure for calculating such important
quantities as the overlap functions and the spectroscopic factors
one can apply it to the one-body density matrices of some open
$s$-$d$ shell nuclei. In order to reveal better the nuclear
structure properties in this region of nuclei and to draw a
parallel with the closed-shell ones, it is desirable to test the
resulting overlap functions in calculations of one-nucleon pickup
and knockout reactions, which is the aim of the present paper.

In the end of this Section we would like to note that the starting
point for the present calculations are OBDM's which include SRC of
central Jastrow type, and which have been computed by a cluster
expansion to leading order in the correlation function. Similar
approach has been very recently used to study the effects of SRC
on the $(e,e^{\prime}p)$ exclusive response functions and cross
sections \cite{Maz2001}. In both calculations the radial
dependence of the correlated function [Eq. (\ref{eq:cf})] is
restricted to Gaussian like one. As well known
\cite{Gai99,Vn97,Fa2001,Ryck2001}, the inclusion of
state-dependent correlations, in particular of tensor ones, is
important when modelling transfer reactions. In addition,
long-range correlations also should be taken into account in a
consistent way. On the other hand, we use in our calculations HO
single-particle wave functions, which do not have the correct
exponential behaviour in the asymptotic region, to construct the
OBDM. It has been already pointed out in previous works
\cite{Iva2001,Maz2001} that an important condition of the
numerical procedure is the exponential asymptotics of the overlap
functions at $r^{\prime}\rightarrow \infty$ [see Eqs.
(\ref{eq:nasym}) and (\ref{eq:pasym})], which is related to the
correct asymptotics of $\rho_{lj}(r,r^{\prime})$ at
$r^{\prime}\rightarrow \infty$. This imposed the development of a
method, though not unique, in \cite{Sto96} to overcome the
difficulties arising from the HO single-particle wave function
asymptotics. The condition for the correct asymptotics is
fulfilled in \cite{Iva2001,Maz2001} by using of a SP basis
obtained with a Woods-Saxon potential. The use of HO wave
functions in the present work instead of more realistic
Woods-Saxon ones following the method from \cite{Sto96} allows us,
however, to obtain analytical expressions for the OBDM's. At
present, no model correlated one-body density matrices treating
open $s$-$d$ shell nuclei are available except that of Ref.
\cite{Mous2000} and, therefore, this work should be considered as
an initial attempt to study the relative importance of the effects
of both SRC and particular structure of these nuclei on overlap
functions, spectroscopic factors and reaction cross sections.

\section{Results of calculations and discussion}

The one-body density matrices (\ref{eq:clusOBDM}) constructed by
using of factor cluster expansion have been applied to calculate
neutron and proton overlap functions including NN correlations and
related to possible $1p$-, $1d$- and $2s$-quasihole states in the
$^{24}$Mg, $^{28}$Si and $^{32}$S nuclei. We note that in our work
the extracted overlap functions are not separated in respect to
the spin-orbit partners $j=l\pm 1/2$. Although the OBDM's do not
allow to obtain different results for $d_{3/2}$ and $d_{5/2}$
quasihole states, it is useful in some calculations of reaction
cross sections for transitions to these states to test the overlap
function corresponding to the $1d$ bound state.

The values of the spectroscopic factors (SF) and of the separation
energies for neutrons and protons deduced from the calculations
are listed in Table 1. The separation energies (\ref{eq:energy})
derived from the procedure are in acceptable agreement with the
corresponding empirical ones. As a common feature, a substantial
reduction of the spectroscopic factors of the states which are
below the Fermi level (of the independent-particle picture) is
observed for all nuclei considered due to the short-range NN
correlations and their particular open-shell structure. For
instance, the values of the SF for the states with $l$=0 and $l$=2
are much smaller than the values obtained for the $l$=1 state. The
particular features of the nuclear structure in the $2s$-$1d$
region (in which the $1p$ state is not a valence one) are
responsible for the larger reduction of the spectroscopic factors
of the levels, though partly occupied, in these open $s$-$d$ shell
nuclei in comparison with the "more occupied" $1p$ quasihole
state. It can be also seen from Table 1 that the trend of the
calculated SF for proton bound states follows that one
corresponding to the neutron overlap functions.

We would like to note the larger reduction of the spectroscopic
factors corresponding to quasihole states of the open-shell nuclei
considered in comparison with the spectroscopic factors for the
states in closed-shell $^{16}$O \cite{Sto96,Gai99} and $^{40}$Ca
\cite{Sto96} nuclei when SRC are taken into account. To discuss
this point, firstly we give in Table 1 the spectroscopic factors
for the $1d$ and $2s$ quasihole states of the closed-shell
$^{40}$Ca deduced from the one-body density matrix calculations
within the same Jastrow approach \cite{Mous2000} used in the
present work. For comparison, the SRC accounted for within the LOA
to the Jastrow correlation method \cite{Sto96} produce
spectroscopic factors of 0.892 for the $1d$ state and 0.956 for
the $2s$ in $^{40}$Ca nucleus. It turns out that the method from
\cite{Mous2000} leads to a suppression of the spectroscopic
factors for the closed-shell $^{40}$Ca nucleus. The reason for the
different SF values could be related to the details of the
calculations with the Jastrow approach when using the two cluster
expansions. Particularly they concern the truncation of the
expansions mentioned already in Section II, which causes a
different account for the SRC (although they are of the same
Jatrow-type), as well as the different asymptotic behaviour of the
OBDM obtained in both expansions. In our opinion, the mentioned
features of the method from \cite{Mous2000} lead to an additional
reduction of the values of the SF for open-shell nuclei as well.
To support this we would like to note the comparison of both
cluster expansions given in \cite{Mous2001} on the example of the
nucleon momentum distributions $n(k)$ of various nuclei. The
high-momentum tails of $n(k)$ obtained using LOA \cite{Sto93}
underestimate the corresponding ones obtained with the use of the
cluster expansion of Clark and co-workers \cite{Mous2000} for the
nuclei considered. In this way the smaller values for the
spectroscopic factors derived in the present work within
\cite{Mous2000} for both closed- and open-shell nuclei reflect the
stronger accounting for the SRC and can be understood. As a common
feature, the spectroscopic factors for the $s$-$d$ open-shell
nuclei are smaller than those for the closed-shell nuclei.

Second, in order to check our theoretically calculated
spectroscopic factors it is useful to compare them with some
single-particle occupation probabilities available for the
considered $1d$-$2s$ shell nuclei. The HF calculations performed
in \cite{Amos78} for the $(p,p^{\prime })$ scattering analyses
treating protons and neutrons equivalently yield values of 0.479
and 0.682 respectively for the $1d_{5/2}$ fractional occupancies
of $^{24}$Mg and $^{28}$Si. The $2s_{1/2}$ proton occupancy of
0.67(9) derived for $^{32}$S was reported in Ref.\cite{Wes97}. As
can be seen from Table 1 our calculated spectroscopic factors for
these states are smaller than the resulting occupancies thus
satisfying the general property $S_{nlj}\leq N_{nlj}^{max}$, i.e.
in each $lj$ subspace the spectroscopic factor $S_{nlj}$ is
smaller than the largest natural occupation number $N_{nlj}^{max}$
\cite{Vn93}.

Generally, the values of the spectroscopic factors for open-shell
nuclei are influenced by the presence of SRC accounted for in our
OBDM similarly to the case of the closed-shell nuclei. However, in
our approach an additional reduction of the spectroscopic factors
for open $1d$-$2s$ shell nuclei takes place because the SF contain
themselves a structure information for the non-complete shell
occupancy in these nuclei. This conclusion is confirmed by the
check of the normalization condition for the OBDM [Eq.
(\ref{eq:ovf})].

The DWBA calculations of the pickup reactions on the considered
open-shell nuclei were performed using the DWUCK4 code \cite{La93}
assuming zero-range approximation for the $p$-$n$ interaction
inside the deuteron. The same approach has been adopted in
previous analyses of $(p,d)$ reactions on $^{16}$O
\cite{Di97,Gai99} and $^{40}$Ca \cite{Di97,Iva2001}. The values of
the optical potential parameters have been taken in each case to
be the same as in the corresponding standard DWBA calculations.

The standard DWUCK4 procedure is performed by calculating the
bound-neutron wave function using the separation energy
prescription (SEP) and different sets of proton and deuteron
optical model parameters. The optical potential is defined to be
\begin{equation}
V_{opt}=-Vf(x_{0})-i\left (W-4W_{D}\frac{d}{dx_{D}}\right
)f(x_{D})- \left (\frac{\hbar}{m_{\pi}c}\right )^{2}V_{s.o.}({\bf
L}.{\bf \sigma}) \frac{1}{r}\frac{d}{dr}f(x_{s.o.})+V_{c},
\label{17}
\end{equation}
where
\begin{equation}
f(x_{i})=[1+exp(x_{i})]^{-1},
\;\;\;\;x_{i}=(r-r_{i}A^{1/3})/a_{i}, \label{18}
\end{equation}
and $V_{c}$ is the Coulomb potential of a uniformly charged sphere
of radius $r_{c}A^{1/3}$. The proton and deuteron optical model
parameters we use are those of K\"{a}llne and Fagerstr\"{o}m
\cite{Kal75} for $^{24}$Mg and $^{32}$S and those of Sundberg and
K\"{a}llne \cite{Sun69} for $^{28}$Si at $E_{p}$=185 MeV incident
energy.

The results for the differential cross sections for the
$^{24}$Mg$(p,d)$, $^{28}$Si$(p,d)$ and $^{32}$S$(p,d)$ reactions
at incident proton energy $E_{p}$=185 MeV compared with the
experimental data are presented in Figs. \ref{fig:pdmg},
\ref{fig:pdsi} and \ref{fig:pds}, respectively. In each panel, the
results obtained with overlap function, with bound-state wave
function following the standard DWBA procedure within the SEP and
with uncorrelated shell-model wave function are plotted. As can be
seen, in general, the use of overlap functions derived from the
one-body density matrix calculations leads to a qualitative
agreement with the experimental data reproducing the amplitude of
the first maximum and qualitatively the shape of the differential
cross section. In this case no extra spectroscopic factor is
needed, since our overlap functions already include the associated
spectroscopic factors. In the examples of some reactions the
standard DWBA form factor is also able to reproduce the shape of
cross sections, while the uncorrelated wave functions fail to
describe correctly neither the size nor the shape of the angular
distributions. The differences between the three types of
calculations are observed mostly at larger angles where the curves
corresponding to the uncorrelated case overestimate substantially
the experimental data. Thus, the relevant importance of SRC
included in our approach becomes clear for a correct description
of pickup processes and this makes it possible to conclude that
the replacement of the overlap integral by a single-particle wave
function is a rough approximation.

We would like to note that our results for the $(p,d)$ cross
sections with the use of overlap functions are obtained without
any additional normalization while the standard DWBA curves and
those corresponding to the uncorrelated case need an adjustment by
a fitting parameter, i.e., the phenomenological spectroscopic
factor. The values of these spectroscopic factors are given in
Table 2 and can be compared with our theoretically calculated
spectroscopic factors from Table 1. For instance, the values of
the phenomenological spectroscopic factors deduced from the
traditional DWBA calculations for the transitions to $5/2^{+}$ and
$3/2^{+}$ in $^{23}$Mg are very close to the empirical values
obtained in \cite{Kal75}, although a clear physical interpretation
could not be done.

Fig. \ref{fig:pdmg} shows the differential cross sections for the
transitions to the ground $3/2^{+}$ state and to the excited
$5/2^{+}$ [at excitation energy $E_{x}$=0.45 MeV] and $1/2^{-}$
[at $E_{x}$=2.76 MeV] states in $^{23}$Mg nucleus. A comparison
with the experimental data from \cite{Kal75} is also made. As can
be seen our calculations using the overlap function for the
transition to the ground $3/2^{+}$ state agree fairly well with
the experimental angular distribution for the same transition
reproducing the amplitude of the first maximum and qualitatively
the shape of the differential cross section. {\it We emphasize
that this result is obtained without any additional normalization
since the calculated overlap function for $3/2^{+}$ state already
includes the spectroscopic factor of 0.4} (see Table 1). Moreover,
at larger angles our calculations lead to better agreement with
the experimental data comparing with the standard DWBA analysis.
In the latter the overlap function is replaced by a SP wave
function corresponding to a given mean-field potential. In such
calculations the NN correlations are included approximately by
adjusting the mean-field potential parameter values. Another
reason for the observed discrepancies in \cite{Kal75} is that a
spherical potential has been used for generating the bound neutron
state in DWBA instead of a more realistic deformed one. We would
like to mention here {\it the principal role of the overlap
function for the good description of the differential cross
section} for the transition to the ground $3/2^{+}$ state in
$^{23}$Mg. It is obtained on the basis of a correlated OBDM and
allows one to account for the short-range correlations in the case
of pickup reaction. The results of the calculations carried out
additionally for the transitions to the excited states are less
satisfactory for angles larger than 10 degrees. One of the reasons
for this is the unrealistic (HO) asymptotic behaviour of the
Jastrow-type OBDM \cite{Mous2000} which has been used to calculate
the overlap functions following the method described in Section
II. Concerning the similarity of the shape of the theoretical
curves (though different by one order of magnitude) in the upper
and middle panels of Fig. \ref{fig:pdmg} we note that for the
ground state the $1d_{3/2}$ overlap function is needed while for
the first excited state the $1d_{5/2}$ overlap function is needed.
The usage of a common $1d$ overlap function in our work obviously
gives a larger deviation from the data for the latter case. On the
other hand, a large fragmentation of the single-particle strengths
is a common feature of the studied $2s$-$1d$ nuclei. For example,
pickup from the $1p_{1/2}$ an 0 $1p_{3/2}$ sub-shells in $2s$-$1d$
nuclei has been an intriguing topic from experimental point of
view \cite{Kal75} indicating several $1p$ states observed in
$^{23}$Mg.

Fig. \ref{fig:pdsi} shows the ground-state differential cross
section of the reaction $^{28}$Si$(p,d)^{27}$Si with $1d_{5/2}$
neutron pickup, while the angular distributions of the reaction
$^{32}$S$(p,d)^{31}$S for the transition to the ground $1/2^{+}$
state and to the excited $3/2^{+}$ [at $E_{x}$=1.24 MeV],
$5/2^{+}$ [at $E_{x}$=2.24 MeV] and $1/2^{-}$ [at $E_{x}$=7.71
MeV] states in $^{31}$S nucleus are shown in Fig. \ref{fig:pds}.
In general, a quantitative agreement of the calculations with the
experimental cross sections in the region of the first maximum is
obtained for all states of the residual $^{27}$Si and $^{31}$S
nuclei. The presence of admixtures of various states observed in
the ground state of $^{32}$S \cite{Kal75} is responsible for a
poorer description of these $(p,d)$ data. The behaviour of the
theoretically calculated angular distribution and the comparison
with the data for the transition to the strongly excited $1/2^{-}$
are similar to those for $l=1$ transfer in
$^{24}$Mg$(p,d)^{23}$Mg.

In order to see the sensitivity of the results to the optical
potential parameters, we examine the transition to the ground
$5/2^{+}$ state in $^{27}$Si nucleus. In Fig. \ref{fig:pdop} three
different theoretical curves are given in respect to deuteron
optical potential parameter values used in the calculations.
Particularly, the effects of changing the radius of the real part
of this potential $R_{d}$ is shown. The best agreement with the
experimental data is achieved with the value of $R_{d}$=0.80 fm
giving also the best DWBA fit in \cite{Sun69}. It can be noted
that the choice of the radius of the real part of the deuteron
optical potential within the interval from 0.65 fm to 0.95 fm
(considered in the standard DWBA analyses) does not influence
strongly the good overall agreement of the cross sections obtained
by means of the theoretically calculated overlap function with the
experimental data. Apart from the shown sensitivity of the
calculations to the deuteron optical potential, in general, we
should mention that the $(p,d)$ reaction is more sensitive to the
reaction mechanism adopted than to the choice of the bound-state
wave function. Nevertheless, it is seen from Fig. \ref{fig:pdop}
that our theoretically calculated overlap function corresponding
to the $1d$ bound state is able to reproduce well the absolute
cross section.

We have analyzed the $(p,d)$ pickup reaction in $^{24}$Mg,
$^{28}$Si and $^{32}$S at proton incident energy of 185 MeV.
Unfortunately, little is known about the deuteron optical
potentials at high energies and the discrepancies with the
experimental data might be partly attributed to the uncertainties
in the determination of the optical potentials. This makes also
difficult to separate clearly the role of the overlap functions
associated with the single-neutron bound states. We emphasize,
however, the important general necessity the overlap functions to
be calculated on the basis of realistic OBDM's and only then the
sensitivity to all other ingredients of the theoretical schemes to
be analyzed.

It turned out from the previous analyzes of one-nucleon removal
reactions \cite{Iva2001,Gai2000} that the quasifree nucleon
knockout is more suitable to investigate the role of overlap
functions as bound-state wave functions. An example of
electron-induced proton knockout from $^{32}$S for the transition
to the ground $2s_{1/2}$ state of $^{31}$P is given in Fig.
\ref{fig:eeps}. Calculations have been done with the code DWEEPY
\cite{DWEEPY}, which is based on the nonrelativistic distorted
wave impulse approximation (DWIA) description of the nucleon
knockout process and includes final-state interactions and Coulomb
distortion of the electron waves \cite{Oxford}. The latter has
been treated with a high-energy expansion in inverse powers of the
electron energy \cite{DWEEPY}. In the figure the result obtained
with the proton overlap function for the $2s$ state of $^{32}$S
and the optical potential from \cite{Schwandt} is compared with
the NIKHEF data from \cite{Wes92}. A reasonable agreement with the
experimental data for the reduced cross section is obtained. In
the analysis of \cite{Wes92} the calculations are performed within
the same DWIA framework and with the same optical potential, but a
phenomenological single-particle wave function is used with a
radius adjusted to the data. {\it We emphasize that in the present
work the overlap function is theoretically calculated on the basis
of the Jastrow-type OBDM of $^{32}$S and does not contain any free
parameters}. It can be seen from Fig. \ref{fig:eeps} that our
spectroscopic factor of 0.5648 gives a good agreement with the
size of the experimental cross section and, in addition, it is in
accordance with the integrated strength for the valence $2s_{1/2}$
shell in $^{32}$S \cite{Wes92} which amounts to 65(7)\% of the SP
strength obtained using the shell-model bound-state function. The
result for the $^{32}$S$(e,e^{\prime}p)$ cross section obtained
with the harmonic-oscillator bound-state wave function is also
illustrated in Fig. \ref{fig:eeps}. It has been computed with the
same oscillator parameter value $b$=2 fm for the $2s_{1/2}$
ground-state wave function as in the original calculations of the
OBDM without SRC \cite{Mous2000}. In order to perform a consistent
comparison with the result when considering theoretically
calculated overlap function, we have applied the same
spectroscopic factor of 0.5648. In this case, the size of the
reduced cross section is also reproduced, but the HO wave function
gives much worse description of the experimental data. Thus, the
comparison made in Fig. \ref{fig:eeps} shows the important role of
the SRC accounted for in our approach for the correct description
of knockout reactions.

We have performed also calculations for the transition to the
excited $1d_{5/2}$ state at $E_{x}$=2.234 MeV and have found a
good agreement with the shape of the experimental momentum
distribution. In our opinion, however, the comparison for the $1d$
states is not very meaningful since our approach cannot
discriminate between the $d_{5/2}$ and $d_{3/2}$ states and the
proton overlap function is calculated for the $1d$ state. From the
experimental point of view, it is noted in \cite{Wes92} that it is
also impossible to distinguish strengths originating from both
$1d$ shells for the integrated $l=2$ strength. In this sense, our
spectroscopic factor of 0.6636 fits well the integrated
spectroscopic strength in the $1d$ shell up to 24 MeV excitation
energy (the value 6.04(60) from Table 5 in \cite{Wes92}) thus
representing a depletion of about 60\% of the full $1d$ shell.

\section{Conclusions}
The results of the present work can be summarized as follows:
\newline
i) Single-particle overlap functions, spectroscopic factors and
separation energies are calculated from the Jastrow-type one-body
density matrices, which were derived using factor cluster
expansion, for the ground state of the open-shell $^{24}$Mg,
$^{28}$Si and $^{32}$S nuclei.
\newline
ii) Taking into account both short-range correlations and specific
nuclear structure it is found that the deduced spectroscopic
factors $S_{nlj}$ of the hole states of open-shell nuclei in our
particular Jastrow approach are substantially smaller than those
of the closed-shell ones satisfying at the same time the general
relation with the natural occupation probabilities $N_{nlj}$,
namely $S_{nlj}\leq N_{nlj}^{max}$.
\newline
iii) The absolute values of the differential cross sections of
$(p,d)$ reactions on $^{24}$Mg, $^{28}$Si and $^{32}$S as well as
of $^{32}$S$(e,e^{\prime}p)$ reduced cross section are calculated
by using the theoretically obtained overlap functions which
already contain NN correlations. The acceptable agreement with the
experimental data shows that this method is applicable also to the
case of open-shell nuclei. The description of the cross sections
might be improved including the deformation effects in the
even-even $Z$=$N$ nuclei in the $2s$-$1d$ region.
\newline
iv) Using more realistic OBDM's with a correct asymptotic
behaviour one can expect that the resulting overlap functions will
be able to describe more accurately the experimental cross
sections of the one-nucleon removal reactions on open $s$-$d$
shell nuclei. In this case more definite conclusions about the
role of the SRC in open- and closed-shell nuclei can be drawn.

\acknowledgments The authors are grateful to Dr. L. Lapik\'{a}s
for providing us the experimental data from \cite{Wes92} and for
the fruitful discussions with him and Prof. K. Amos. This work was
partly supported by the Bulgarian National Science Foundation
under the Contracts Nrs.$\Phi $--809 and $\Phi $--905.

\newpage
\noindent {\bf Table 1:} Neutron and proton spectroscopic factors
(SF) and separation energies ($\epsilon_{n}$ and $\epsilon_{p}$ in
MeV) calculated on the basis of the one-body density matrices
\cite{Mous2000} for $^{24}$Mg, $^{28}$Si, $^{32}$S and $^{40}$Ca.
Comparison is made with the experimental data $\epsilon_{n}^{exp}$
and $\epsilon_{p}^{exp}$ for the separation energies
\cite{Kal75,Sun69,Wes92,End67,Mah91}.

\begin{center}
\begin{tabular}{cccccccccccc}
\hline\hline
& & & \multicolumn{3}{c}{neutrons} & &  \multicolumn{3}{c}{protons} \\
\cline{4-6} \cline{8-10} Nucleus & $nl$ & & $\epsilon_{n}$ &
$\epsilon_{n}^{exp}$ & SF  & & $\epsilon_{p}$ &
$\epsilon_{p}^{exp}$ & SF \\
\hline

$^{24}$Mg  & $1p$  & & 18.85 & 19.29 & 0.9474 & & 14.13 & 14.33 & 0.9798\\
           & $1d$  & & 16.94 & 16.53 & 0.4026 & & 11.02 & 11.69 & 0.4586\\
           & $2s$  & & 18.37 & 18.89 & 0.4706 & & 13.63 & 14.08 & 0.5414\\
\hline
$^{28}$Si  & $1p$  & & 21.34 & 21.35 & 0.8528 & & 15.21 & 15.64 & 0.9788\\
           & $1d$  & & 17.29 & 17.18 & 0.5071 & & 11.49 & 11.59 & 0.6265\\
           & $2s$  & & 17.32 & 17.96 & 0.3958 & & 12.09 & 12.43 & 0.4675\\
\hline
$^{32}$S   & $1p$  & & 22.54 & 22.80 & 0.7454 & & 15.68 & 16.29 & 0.8856\\
           & $1d$  & & 16.98 & 17.33 & 0.5682 & &  9.82 & 10.13 & 0.6636\\
           & $2s$  & & 14.75 & 15.09 & 0.4712 & &  8.53 &  8.86 & 0.5648\\
\hline
$^{40}$Ca  & $1d$  & & 15.45 & 15.64 & 0.7691 & &  8.34 &  8.33 & 0.6729\\
           & $2s$  & & 17.86 & 18.19 & 0.8001 & & 10.33 & 10.94 & 0.8051\\
\hline\hline
\end{tabular}
\end{center}

\vspace{3cm} \noindent {\bf Table 2:} Phenomenological
spectroscopic factors $S_{DWBA}$ deduced from the standard DWBA
calculations and $S_{SM}$ deduced from the calculations with
uncorrelated shell-model (SM) OBDM's of Ref. \cite{Mous2000} for
the reactions considered in Figs. \ref{fig:pdmg}-\ref{fig:pds}.

\begin{center}
\begin{tabular}{cccccc}
\hline\hline
Reaction & & $nl_{j}$ & $S_{DWBA}$ & $S_{SM}$ \\
\hline

$^{24}$Mg$(p,d)$$^{23}$Mg  & & $1d_{3/2}$ & 0.25 & 0.10 \\
                           & & $1d_{5/2}$ & 2.50 & 1.00 \\
                           & & $1p_{1/2}$ & 1.11 & 0.75 \\
\hline
$^{28}$Si$(p,d)$$^{27}$Si  & & $1d_{5/2}$ & 2.13 & 2.13 \\
\hline
$^{32}$S$(p,d)$$^{31}$S    & & $2s_{1/2}$ & 0.33 & 0.56 \\
                           & & $1d_{3/2}$ & 1.43 & 1.00 \\
                           & & $1d_{5/2}$ & 4.00 & 1.00 \\
                           & & $1p_{1/2}$ & 1.00 & 0.11 \\
\hline\hline
\end{tabular}
\end{center}

\newpage
\begin{figure}
\caption{Differential cross-section for the $^{24}$Mg$(p,d)$
reaction at incident proton energy $E_{p}$=185 MeV to the
$3/2^{+}$ ground and to the $5/2^{+}$ and $1/2^{-}$ excited states
in $^{23}$Mg. The neutron overlap functions are derived from the
OBDM (solid line). The calculated standard DWBA curve (dotted
line) and the uncorrelated SM result (dashed line) are also
presented. The experimental data \cite{Kal75} are given by the
full circles. \label{fig:pdmg}}
\end{figure}

\begin{figure}
\caption{Differential cross-section for the $^{28}$Si$(p,d)$
reaction at incident proton energy $E_{p}$=185 MeV to the
$5/2^{+}$ ground state in $^{27}$Si. The neutron overlap functions
are derived from the OBDM (solid line). The calculated standard
DWBA curve (dotted line) and the uncorrelated SM result (dashed
line) are also presented. The experimental data \cite{Sun69} are
given by the full circles. \label{fig:pdsi}}
\end{figure}

\begin{figure}
\caption{Differential cross-section for the $^{32}$S$(p,d)$
reaction at incident proton energy $E_{p}$=185 MeV to the
$1/2^{+}$ ground and to the $3/2^{+}$, $5/2^{+}$ and $1/2^{-}$
excited states in $^{31}$S. The neutron overlap functions are
derived from the OBDM (solid line). The calculated standard DWBA
curve (dotted line) and the uncorrelated SM result (dashed line)
are also presented. The experimental data \cite{Kal75} are given
by the full circles. \label{fig:pds}}
\end{figure}

\begin{figure}
\caption{Differential cross-section for the $^{28}$Si$(p,d)$
reaction at incident proton energy $E_{p}$=185 MeV to the
$5/2^{+}$ ground state in $^{27}$Si. Line convention referring to
calculations using different optical potential parameter values
and neutron overlap function derived from the OBDM is given (see
also the text). The experimental data \cite{Sun69} are given by
the full circles. \label{fig:pdop}}
\end{figure}

\begin{figure}
\caption{Reduced cross section of the $^{32}$S($e,e'p$) reaction
as a function of the missing momentum $p_{\mathrm m}$ for the
transition to the $1/2^{+}$ ground state of $^{31}$P. The proton
overlap function is derived from the OBDM (solid line). The result
with the uncorrelated (HO) wave function is given by dotted line.
The experimental data (full circles) are taken from Ref.
\cite{Wes92}. \label{fig:eeps}}
\end{figure}

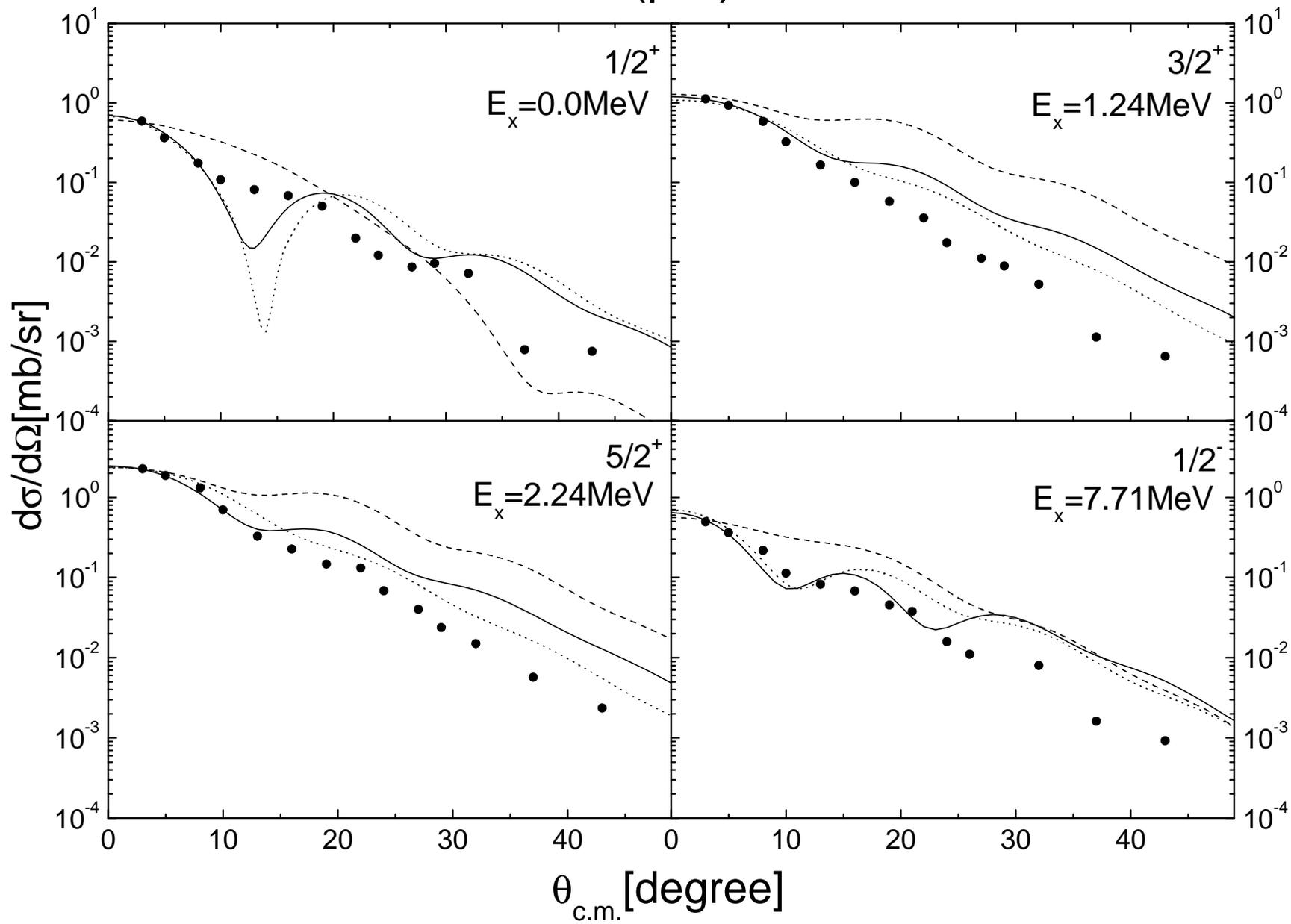
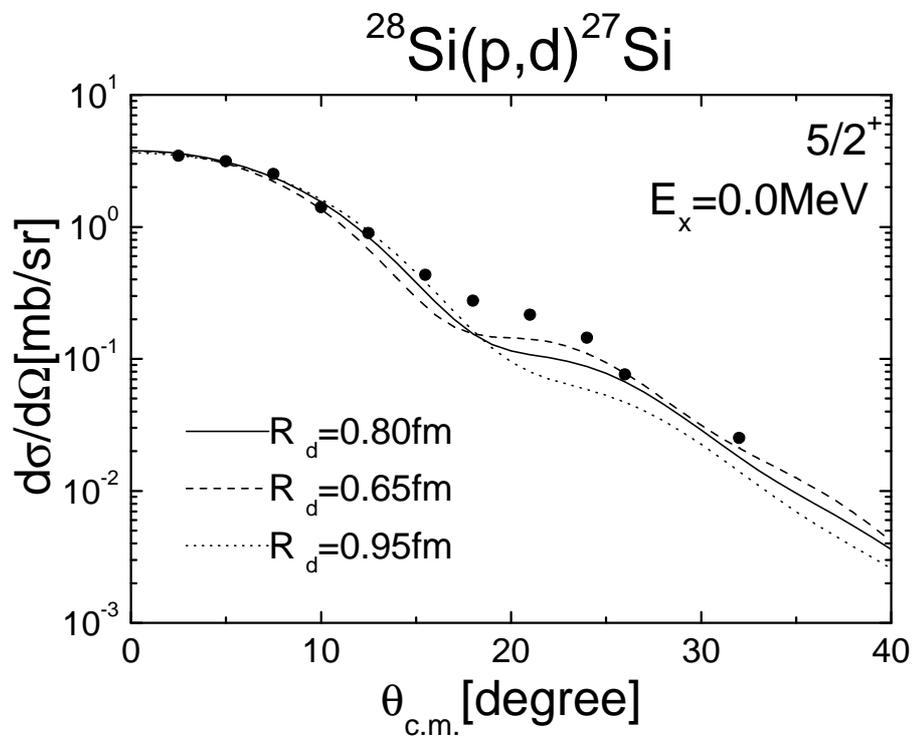
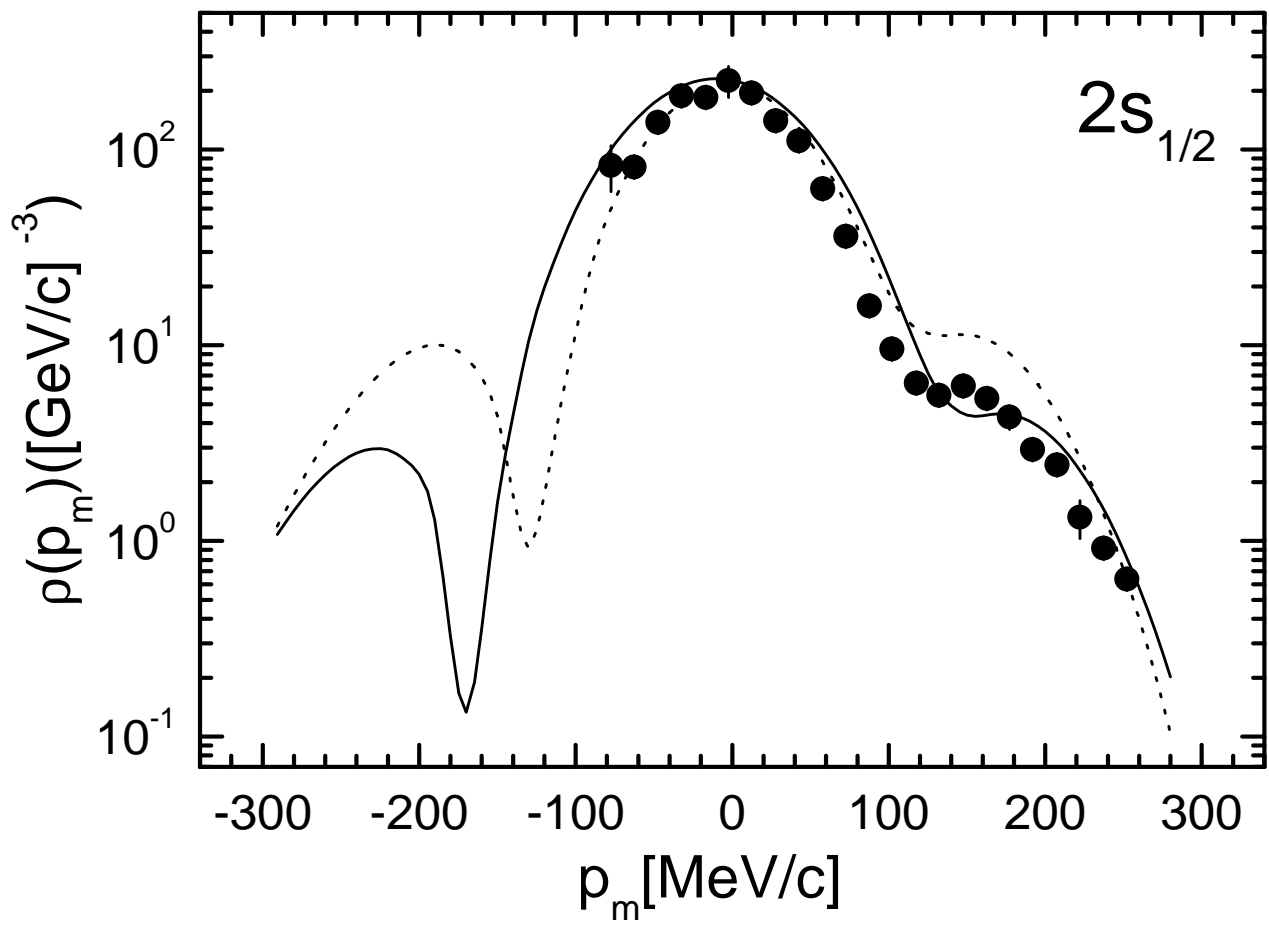
\end{document}